# Paving the way to carbon neutrality: Evaluating the decarbonization of residential building electrification worldwide☆

Yuanyuan Wang [1†], Minda Ma [2*, 3,†], Nan Zhou [3], Zhili Ma [1]

1. School of Management Science and Real Estate, Chongqing University, Chongqing, 400045, PR China
2. School of Architecture and Urban Planning, Chongqing University, Chongqing, 400045, PR China
3. Building Technology and Urban Systems Division, Energy Technologies Area, Lawrence Berkeley National Laboratory, Berkeley, CA 94720, United States

- Corresponding author: Dr. Minda Ma, Email: maminda@lbl.gov
  Homepage: https://buildings.lbl.gov/people/minda-ma
  https://chongjian.cqu.edu.cn/info/1556/6706.htm

---

☆ The coauthors from Lawrence Berkeley National Laboratory declare that this manuscript was authored by an author at Lawrence Berkeley National Laboratory under Contract No. DE-AC02-05CH11231 with the U.S. Department of Energy. The U.S. Government retains, and the publisher, by accepting the article for publication, acknowledges, that the U.S. Government retains a non-exclusive, paid-up, irrevocable, world-wide license to publish or reproduce the published form of this manuscript, or allows others to do so, for U.S. Government purposes.

†: The two authors contributed equally to the study.

## Graphical abstract

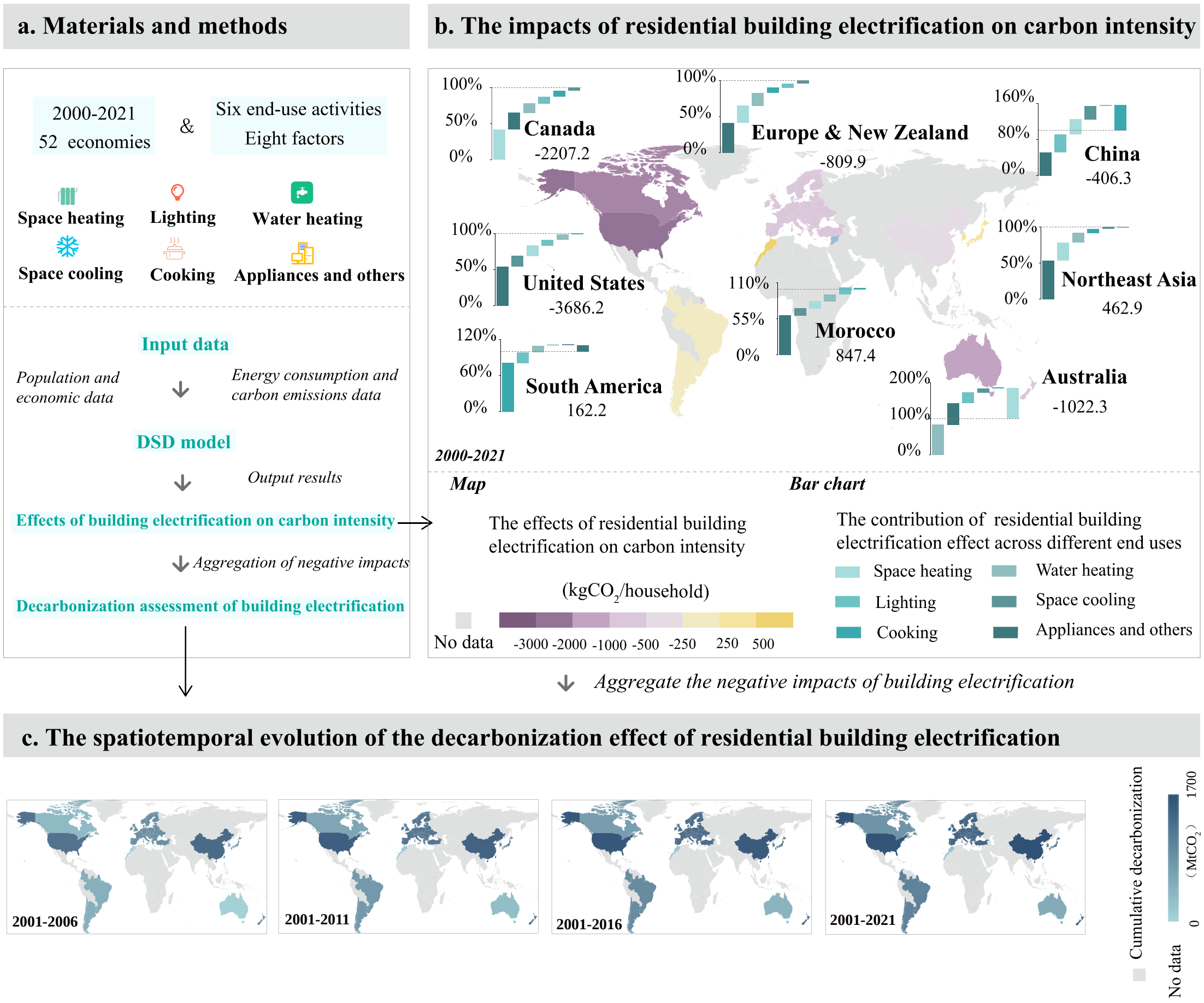

**Graphical abstract.** The decarbonization impacts of electrification of residential buildings worldwide: (a) the research framework; (b) the impacts of residential building electrification on carbon intensity; (c) the spatiotemporal evolution of the decarbonization effect of residential building electrification. Note: Considering the data availability, lighting was excluded in Northeast Asia.

## Highlights

- Global residential electrification rose from 29.9% in 2000 to 40.1% in 2021.
- Global residential electricity emissions rose from 1452 to 2032 Mt$CO_2$ in 2000–2021.
- Electrification does not always lead to decarbonization in residential buildings.
- Residential electrification achieved decarbonization of 188 Mt$CO_2$ per yr worldwide.
- Power decarbonization is key to reducing emissions through electrification.

## Abstract

In the context of increasing global climate change, decarbonizing the residential building sector is crucial for sustainable development. This study aims to analyze the role of various influencing factors in carbon intensity changes using the decomposing structural decomposition (DSD) to assess and compare the potential and effectiveness of electrifying end-use activities during the operational phase of residential buildings worldwide for decarbonization. The results show that (1) while the electrification rate varied in its impact on emissions across different countries and regions, the overall increase in electrification contributed to higher carbon intensity. In contrast, changes in the emission factor of electricity generally made a positive contribution to emission reduction globally. (2) The global electrification level has significantly increased, with the electrification rate rising from 29.9% in 2000 to 40.1% in 2021. A 39.8% increase in the electricity-related carbon emissions of global residential buildings was observed, increasing from 1452 $MtCO_2$ to 2032 $MtCO_2$, 2000–2021. (3) From 2000 to 2021, electrification of space heating was the main contributor to carbon reduction, whereas the contributions of electrification to cooling and lighting were relatively limited. Emission reductions from appliances and others remained stable. The electrification of water heating and cooking had varying effects on emission reductions in different countries. Furthermore, this study proposes a series of electrification decarbonization strategies. Overall, this study analyzes and contrasts decarbonization efforts from building electrification at the global and regional levels, explores the key motivations behind these efforts to aid national net-zero emission targets and accelerate the transition of the global residential building sector toward a carbon-neutral future.

## Abbreviation notation

DSD – Decomposing structural decomposition

GDP – Gross domestic product

HFC – Household final consumption

$kgCO_2$ – Kilograms of carbon dioxide

$MtCO_2$ – Mega tons of carbon dioxide

## Nomenclature

$C$ – Total carbon emissions during residential building operations

$c$ – Carbon emissions released by each household

$C_i$ $(i = 1, \ldots, 6)$ – Carbon emissions from six end uses

$C_{iele}$ $(i = 1, \ldots, 6)$ – Carbon emissions related electricity use from six end uses

$DC$ – Total decarbonization

$DCI$ – Decarbonization per household

$E$ – Energy consumption during residential building operations

$E_i$ $(i = 1, \ldots, 6)$ – Energy consumption by end-use $i$

$E_{iele}$ $(i = 1, \ldots, 6)$ – Energy consumption related electricity use from six end uses

$e$ – Energy intensity

$g$ – GDP per capita

$H$ – Amount of households

$h$ – Household consumption capacity

$k$ – Electricity emission factor

$m$ – Electrification rate

$n$ – The factor of total emissions relative to electricity-related carbon emissions in each end-use activity (expression is $\frac{C_i}{C_{iele}}$)

$P$ – Population

$p$ – Average household size

$s$ – End-use structure

## 1. Introduction

The building sector plays a major role in worldwide energy consumption [1, 2], and its decarbonization is vital amid worsening climate change [3]. Residential buildings, a major sub-sector, consume substantial energy and produce large amounts of carbon dioxide during operations [4], particularly from end uses such as space heating [5, 6], space cooling, lighting, and cooking [7]. In 2022, the International Energy Agency (IEA) reported that residential buildings accounted for about 25% of global energy consumption [8], driven by increasing urbanization and housing demand [9]. Electrification is emerging as a key strategy for low-carbon transition in buildings [10], focusing on converting energy end uses—such as space heating and water heating [11]—into electricity-driven systems. This shift reduces reliance on fossil fuels and enhances energy efficiency [12]. As a result, residential building electrification is widely regarded as an effective pathway for reducing carbon emissions.

End-use electrification is rapidly becoming a central trend in the evolution of building energy systems [13]. According to the IEA, expanding electrification—alongside a cleaner electricity mix—is critical for the transition to a low-carbon energy future. One key measure in this process is the replacement of coal-fired boilers with electric heat pumps, which significantly reduces direct emissions in buildings [14]. Beyond this, electrifying gas-based systems—including space heating, water heating, and various household appliances—is also essential to accelerate the decarbonization of residential buildings [15]. As electrification advances, residential electricity demand has grown, driven by increasing reliance on electric appliances and the adoption of smart home technologies [16]. Despite this growing importance, most existing studies on building electrification remain focused on specific countries or isolated end-use activities. Furthermore, traditional analytical methods have clear limitations: the Logarithmic Mean Divisia Index (LMDI) often overlooks interactions among influencing factors, while the Generalized Divisia Index Method (GDIM) fails to account for the roles of specific end-use activities [17]. Addressing these gaps, there remains a lack of comprehensive global studies that assess residential building electrification, particularly those offering detailed decarbonization insights across various end-use activities. Thus, this study proposes three key questions regarding residential building electrification during the operational phase:

- How does the end-use electrification impact the carbon intensity of global residential buildings?
- What progress has been made in the electrification of residential buildings since 2000 worldwide?
- How does end-use electrification contribute to decarbonization, and how can its progress be promoted?

To answer the questions posed above, a bottom-up model is developed to measure the influence of electrification of household end-use activities in residential buildings on carbon emission intensity during the operational phase across 52 countries in 11 major emission regions from 2000 to 2021. The objective of this study is to apply the decomposing structural decomposition (DSD) method to establish a model that quantifies the effects of electrifying diverse end-use activities on lowering carbon emissions. Besides, this study examines electrification rates and emission factors of electricity, analyzing and comparing the decarbonization impacts across various end-use activities. Additionally, the study offers recommendations for accelerating residential building electrification and advancing power decarbonization.

**The key contribution of this study** lies in the development of a robust evaluation model to quantify the historical carbon emission intensity of global residential buildings and to assess the role of building electrification in the decarbonization process. Building upon the original DSD method [18], this study extends the method by incorporating factors specifically related to residential building electrification. Furthermore, it explores the evolution of decarbonization pathways associated with electrification across three critical dimensions: temporal trends, regional heterogeneity and end-use activities. The results provide not only references to support the contribution of residential building electrification to global carbon emissions reduction and decarbonization potential but also valuable guidance for policymakers in shaping practical and impactful strategies.

The remains of this study are arranged as follows: Section 2 contains a review of the literature. Section 3 describes the methodology, focusing on the construction of the carbon emission model, the DSD method and the decarbonization assessment model. Section 4 reveals the impact of residential building electrification on carbon intensity during the operational phase. Section 5 discusses the historical decarbonization evaluation of electrification and outlines some decarbonization strategies. Finally, Section 6 highlights the significant results of this study and offers suggestions for future research.

## 2. Literature review

In carbon emission studies related to the building sector, decomposition analysis techniques are frequently applied to examine how various factors affect emission changes [19]. The LMDI, an established method, allocates variations in carbon emissions to various influencing factors [20]. LMDI is a widely accepted method for assessing carbon dioxide ($CO_2$) emission reduction because of its ease of calculation and lack of residuals [21]. However, its limitation lies in addressing the interdependence among variables, which may result in interference with the analysis outcomes due to relationships between factors. In addition, LMDI has limitations in integrating multiple absolute and relative factors, which limits its application in complex analyses [22]. The GDIM is introduced to address this limitation. The GDIM, which is built on LMDI, features a flexible design that better represents the factors affecting carbon emissions, excelling particularly in analyzing long-term trends [23]. Although GDIM has significantly improved the accuracy and independence of decomposition analysis in many cases, it is insufficient for comprehensively analyzing historical carbon emission changes from the point of view of end-use activities [24]. To this end, Boratyński [18] proposed the DSD method, a simplified and intuitive decomposition tool designed to enhance the operability of analysis. The DSD method not only effectively reduces inter-factor interference but also delivers clear results, helping to pinpoint the main contributors to carbon emissions in residential buildings [25].

Residential building electrification is widely regarded as a key pathway for achieving low-carbon transitions [26]. Studies have shown that electrification can significantly reduce carbon emissions, especially when combined with renewable energy sources [27, 28]. Some scholars developed a bottom-up model to assess residential electrification in Italy, demonstrating improvements in energy efficiency and significant reductions in carbon emissions [29]. Bistline et al. [28] assessed building electrification in the United States, highlighting its strong potential to reduce greenhouse gas emissions. Studies on residential heat pump deployment have indicated a potential reduction in $CO_2$ emissions of 38–53%, highlighting the role of electrification in decarbonization [30]. At the urban level, Costanzo, Nocera, Detommaso, and Evola [31] explored the collaborative implementation of electrification and building retrofit measures reduced carbon emissions by 70% in densely built residential areas in Catania.

As building electrification advanced, the global electricity demand has increased significantly [32]. While it increases reliance on electricity and may impose additional stress on power systems during peak periods [27], it also offers opportunities for the integration of renewable energy sources [33]. A case study demonstrated that widespread residential electrification can drive renewable energy usage [34]. Some studies have noted that successfully achieving building electrification requires integration with energy efficiency improvements, renewable energy development [35], and smart grid infrastructure [36]. Vaishnav and Fatimah [37] argued that the effectiveness of electrification largely depends on the structure of the power supply. Studies quantifying end-use electrification remain limited [38], and there are conflicting evaluations of its decarbonization potential [39-41]. The carbon reduction potential of electrification depends on factors such as regional energy mix [42], and the limited efficiency and feasibility of technologies like heat pumps and air conditioners [43], particularly under conditions of seasonal demand [44]. Moreover, user behavior may also hinder expected emission reductions in emissions. Lagging policies and incentives also limit the advancement of building electrification, and many studies lack comprehensive consideration of these factors [45].

In summary, from a methodological perspective, Table 1 compares commonly used decomposition approaches. While the LMDI method is widely applied, its results can be distorted by interdependencies among variables. Similarly, the GDIM approach fails to capture emission changes at the end-use level. As shown in Table 2, much of the existing literature focus on individual countries or regions, which limits the generalizability of the findings. Furthermore, Table 3 shows that many studies restrict their analysis to a single end-use activity, thereby overlooking the broader implications of electrification across various end uses. Therefore, two key issues should be considered when assessing carbon emission reductions through residential building electrification:

**First, regarding the assessment of carbon reduction contributions from residential building electrification,** existing studies have explored various drivers of carbon emissions in the building sector but have paid limited attention to the specific role of electrification in emission reductions. Moreover, traditional decomposition methods remain limited in accurately capturing the contribution of end-use electrification to overall reductions. Since a detailed end-use breakdown allows for more precise decomposition results, the DSD model is well suited to address this analytical gap.

**Second, in terms of the scope of existing studies,** most studies on residential building electrification focus on specific regions or individual end-use activities [46]. There is a lack of studies examining the effects of electrification across different climatic conditions and among countries or regions at varying levels of development. Cross-country comparisons of residential building carbon emissions and energy consumption are essential for evaluating the current status of the building sector, identifying trends in energy efficiency and low-carbon development, and informing strategies to achieve carbon neutrality.

To address the identified gaps, this study employs the DSD approach to assess the progress of the global residential building sector's carbon emission reduction, analyze how residential building electrification affects carbon emissions in different countries, and evaluate the decarbonization progress in global residential building electrification. This study's primary contributions are as follows:

- **This study utilizes the DSD approach to analyze the factors influencing carbon intensity changes resulting from global residential building electrification during operations.** This study assesses the influence of end-use electrification on carbon intensity during the operational phase, emphasizing variations across different end-use activities. The analysis covers six major end-use activities (including cooking, lighting, water heating, appliance and others, space heating, and space cooling). The aim is to provide references for future electrification pathway choices by comparing the specific emission reduction performance of electrification across these end-use activities.
- **This study develops a model to assess the decarbonization impact of electrification on global residential buildings and examines its direct effect on carbon emission reduction.** Through comparative analysis, the study highlights the specific opportunities and challenges faced by different regions in achieving building decarbonization goals, particularly in relation to the decarbonization capabilities and outcomes influenced by variations in policy support, energy structures, and technology application. This study aims to provide valuable references for global pathways to decarbonize building electrification and to assist in formulating more targeted and feasible policies and measures.

**Table 1**. The comparison of the decomposition analysis methods applied in the building sector.

| Reference | Year | Focus | Method | Methodological limitations or contributions |
| --- | --- | --- | --- | --- |
| [20] | 2011 | Decomposition analysis of residential energy consumption in Hong Kong | LMDI | Insufficient consideration of factor-specific relevance in energy consumption patterns |
| [21] | 2023 | Spatiotemporal decomposition of $CO_2$ emissions in China's commercial buildings | LMDI | Lack of detailed analysis at the end-use level (e.g., heating, lighting) |
| [22] | 2020 | Decomposition of global energy cost drivers | LMDI | Insufficient consideration of non-linear relationships among emission drivers |
| [23] | 2014 | Methodological extension of $CO_2$ decomposition approaches | GDIM | Assumption of exponential trends for all quantitative indicators<br>Lack of rigorous validation of methodological robustness |
| [24] | 2023 | Decomposition of carbon emission drivers in China | GDIM | Narrow selection of carbon emission drivers<br>Methodological limitation in end-use level decomposition of emission changes |
| [18] | 2021 | Impacts of structural changes on energy consumption patterns | DSD | Introduction of a novel structural decomposition framework |
| [25] | 2023 | Global trends in operational carbon intensity of residential buildings | DSD | Development of an end-use activity decomposition methodology |

**Table 2**. The summary of studies on carbon reduction through building electrification.

| Reference | Year | Focus | Study region | Problems, challenges, and limitations or gaps of science | Recommendations or findings |
|---|---|---|---|---|---|
| [26] | 2019 | Decarbonization through electrification and renewable integration | China | Requirement of alignment between electrification pace and grid decarbonization | Accelerated electrification and renewables may help China peak $CO_2$ emissions by 2025 |
| [27] | 2021 | Impacts of residential space heating electrification on energy consumption | The United States | Limited to a single end-use activity electrification. | 4.1% emission reduction from standard-efficiency heat pumps |
| [28] | 2022 | Macroeconomic impacts of electrification on $CO_2$ and air pollutants | The Unites States | Insufficient consideration of climate change effects on emission reduction | Significant emission reduction from electrification |
| [29] | 2020 | Assessment of electrification pathways for residential energy consumption | Italy | Limited to distributed heating systems | 56% reduction in primary energy consumption from heat pumps in heating sector |
| [30] | 2022 | Impacts of residential space heating electrification | The Unites States | Simplified assumption of a linear trajectory in grid decarbonization | Potential emission reduction from electrification by 2050 across most regions |
| [31] | 2024 | Urban decarbonization through electrification and retrofitting | Catania, Italy | Limited applicability to colder regions | Significant urban emissions reduction from combined heat pumps and photovoltaics |
| [32] | 2020 | Trade-offs between electrification policies and carbon mitigation | China | Overly optimistic assumptions about renewable energy adoption | 28% emission reduction potential from 80% renewable electricity share |
| [33] | 2024 | Challenges of sustainable electrification | Africa | Unavailability of electricity data for certain regions | Electrification constraints in Africa due to insufficient infrastructure |
| [34] | 2023 | Impacts of building electrification on carbon emissions | California, the United States | Insufficient consideration of regional heterogeneity in electricity generation mix | Substantial carbon emission reduction from building electrification |
| [35] | 2023 | Decarbonization pathways for regional grids under policy scenarios | China | Lack of analysis on inter-regional coordination | Decarbonization potential of government planning scenarios |
| [36] | 2024 | Key issues and challenges of building electrification | The North America | Limited geographically scope of analysis | Feasibility of building electrification depends on infrastructure improvements |

**Table 3**. The comparison of studies on the building end-use activity electrification.

| Reference | Year | Focus | End uses | Problems, challenges, and limitations or gaps of science | Recommendations or findings |
|---|---|---|---|---|---|
| [37] | 2020 | Environmental impacts of space heating electrification in the United States | Space heating | Limited generalizability of findings to other countries | Effective greenhouse gas emission reduction through electrification |
| [38] | 2024 | Coordination of building electrification with regional grid decarbonization in China | Space heating, water heating, and cooking | Lack of updated end-use energy consumption data | Emission reduction in most regions from high-efficiency appliances |
| [39] | 2022 | Challenges of residential space heating electrification in China | Space heating | Restriction of analysis to case studies in two cities | Barriers to residential heating electrification (technology/economics/policy) |
| [40] | 2022 | Electrification of residential heat demand in Italy | Space heating | Insufficient consideration of regional grid constraints | 15% peak electricity demand reduction from heat pumps |
| [41] | 2022 | Load smoothing potential of hybrid systems in Australia | Heating | Limited focus on detached houses (excluding high-rise flats) | Rapid building decarbonization of electrification (heat pumps and photovoltaics) |
| [42] | 2024 | Decarbonization efficiency through electrification in China | All | Reliance on average electricity carbon emission drivers (without marginal factor consideration) | Barrier to emission reduction from slow electricity decarbonization |
| [43] | 2018 | Air quality/health co-benefits of building electrification in China | Space heating and cooling | Insufficient consideration of urban-rural disparities in electrification access | 15% $CO_2$ emission reduction corresponding to 52% lower coal share in electricity generation |

## 3. Materials and methods

This study developed a bottom-up emissions model, integrated with the DSD approach, to analyze the global carbon intensity of residential building operations, with a particular focus on the impact of electrification on carbon reduction in building operations. The research framework is illustrated in Fig. 1.

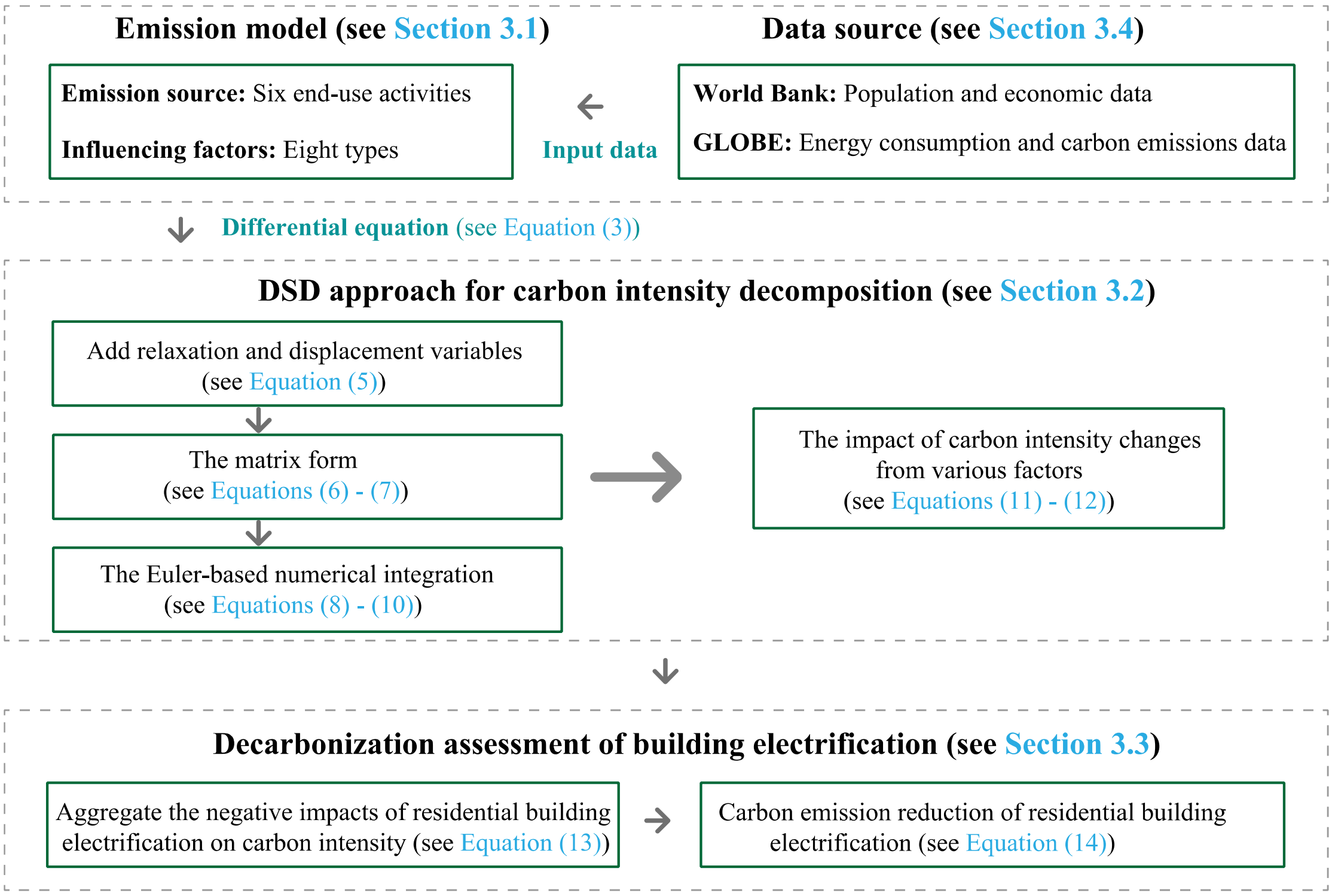

**Fig. 1.** The research framework of this paper.

### *3.1. Residential building operational emissions model*

In residential buildings, carbon emissions during the operational phase were largely due to the energy required for daily household activities. These emissions can be classified according to different functional needs [47]. These are mainly composed of six parts: lighting, cooking, space cooling, appliances and others, space heating and water heating [48].

Thus, the carbon emission model used to calculate global carbon emissions is expressed as follows:

$$C = C_{\text{Water heating}} + C_{\text{Space heating}} + C_{\text{Cooking}} + C_{\text{Lighting}} + C_{\text{Space cooling}} + C_{\text{Appliances \& others}}$$
$$\text{Simplified as} C = \sum_{i=1}^{6} C_i \quad (1)$$

The symbol $C_i$ $(i = 1,2,3,4,5,6)$ denotes the carbon emissions linked to various end-use activities. The emissions of each family household indicate the residential building carbon intensity, expressed as $c_i = \frac{C_i}{H}$ ($H$ represents the number of households).

This study examined eight determinants affecting carbon emission intensity in residential building operations. These include average family size ($\frac{P}{H}$), per capita ($\frac{GDP}{P}$) gross domestic product (GDP), and the consumption capacity of family households ($\frac{HFC}{GDP}$), with HFC denoting household final consumption. The study evaluated additional factors, including energy intensity ($\frac{E}{HFC}$), the structure of end-use activity ($\frac{E_i}{E}$, where $E_i$ is the energy demand associated with specific end-use activity), the electrification rate ($\frac{E_{iele}}{E_i}$), the emission factor of electricity ($\frac{C_{iele}}{E_{iele}}$) and the factor of total emissions relative to electricity-related carbon emissions in each end-use activity ($\frac{C_i}{C_{iele}}$).

$$c_i = \frac{C_i}{H} = \frac{P}{H} \cdot \frac{GDP}{P} \cdot \frac{HFC}{GDP} \cdot \frac{E}{HFC} \cdot \frac{E_i}{E} \cdot \frac{E_{iele}}{E_i} \cdot \frac{C_{iele}}{E_{iele}} \cdot \frac{C_i}{C_{iele}}$$
$$\text{Shorted as} c_i = p \cdot g \cdot h \cdot e \cdot s_i \cdot m_i \cdot k_i \cdot n_i \tag{2}$$

Accordingly, we defined the operational carbon emission model in the following manner:

$$c = \sum_{i=1}^{6} p \cdot g \cdot h \cdot e \cdot m_i \cdot n_i \cdot k_i \cdot s_i \tag{3}$$

*3.2. DSD approach for carbon intensity decomposition*

According to the principle of the DSD method [18], the total differential equation of Equation (3) can be expressed as follows:

$$\mathrm{d}c = \frac{\partial c}{\partial p}\mathrm{d}p + \frac{\partial c}{\partial g}\mathrm{d}g + \frac{\partial c}{\partial e}\mathrm{d}h + \frac{\partial c}{\partial e}\mathrm{d}e + \sum_{i=1}^{6}\left(\frac{\partial c_i}{\partial m_i}\mathrm{d}m_i + \frac{\partial c_i}{\partial n_i}\mathrm{d}n_i + \frac{\partial c_i}{\partial k_i}\mathrm{d}k_i + \frac{\partial c_i}{\partial e_i}\mathrm{d}s_i\right) \tag{4}$$

Then, on the basis of Equation (4), the relaxation variables $\mathrm{d}F_i$ and the displacement variables $\mathrm{d}F$ were added to form a linear equation system:

$$\begin{cases} \mathrm{d}c = \dfrac{\partial c}{\partial p}\mathrm{d}p + \dfrac{\partial c}{\partial g}\mathrm{d}g + \dfrac{\partial c}{\partial e}\mathrm{d}h + \dfrac{\partial c}{\partial e}\mathrm{d}e + \displaystyle\sum_{i=1}^{6}\left(\dfrac{\partial c_i}{\partial m_i}\mathrm{d}m_i + \dfrac{\partial c_i}{\partial n_i}\mathrm{d}n_i + \dfrac{\partial c_i}{\partial k_i}\mathrm{d}k_i + \dfrac{\partial c_i}{\partial e_i}\mathrm{d}s_i\right) \\ \mathrm{d}s_i = \mathrm{d}F_i + \mathrm{d}F \\ \displaystyle\sum_{i=1}^{6}\mathrm{d}s_i = 0 \end{cases} \tag{5}$$

The matrix form of Equation (5) was simplified to the following expression:

$$\mathrm{A} \cdot \mathrm{dy} = \mathrm{B} \cdot \mathrm{dz} \tag{6}$$

dy and dz in Equation (6) represent the endogenous vectors and exogenous vectors, respectively. These vectors were defined as $\mathrm{dy} = [\mathrm{d}c,\mathrm{d}s_1,\mathrm{d}s_2,\cdots,\mathrm{d}s_6,\mathrm{d}F]^{\mathrm{T}}$ and $\mathrm{dz} = [\mathrm{d}p,\mathrm{d}g,\mathrm{d}h,\mathrm{de},\mathrm{d}m_1,\cdots,\mathrm{d}m_6,\mathrm{d}n_1,\cdots,\mathrm{d}n_6,\mathrm{d}k_1,\cdots,\mathrm{d}k_6,\mathrm{d}F_1,\cdots,\mathrm{d}F_6]^{\mathrm{T}}$. Here, A and B represent the matrices of the coefficient associated only with the variables dy and dz, where $\mathrm{A} = \lambda(\mathrm{y},\mathrm{z})$ and $\mathrm{B} = \omega(\mathrm{y},\mathrm{z})$, satisfying the following conditions:

$$\mathrm{A} = \begin{pmatrix} 1 & -\frac{\partial c_1}{\partial s_1} & -\frac{\partial c_2}{\partial s_2} & -\frac{\partial c_3}{\partial s_3} & -\frac{\partial c_4}{\partial s_4} & -\frac{\partial c_5}{\partial s_5} & -\frac{\partial c_6}{\partial s_6} & 0 \\ 0 & 1 & 0 & 0 & 0 & 0 & 0 & -1 \\ 0 & 0 & 1 & 0 & 0 & 0 & 0 & -1 \\ 0 & 0 & 0 & 1 & 0 & 0 & 0 & -1 \\ 0 & 0 & 0 & 0 & 1 & 0 & 0 & -1 \\ 0 & 0 & 0 & 0 & 0 & 1 & 0 & -1 \\ 0 & 0 & 0 & 0 & 0 & 0 & 1 & -1 \\ 0 & 1 & 1 & 1 & 1 & 1 & 1 & 0 \end{pmatrix}$$
$$\mathrm{B} = \begin{pmatrix} \frac{\partial c}{\partial p} & \frac{\partial c}{\partial g} & \frac{\partial c}{\partial h} & \frac{\partial c}{\partial e} & \frac{\partial c_1}{\partial m_1} & \cdots & \frac{\partial c_6}{\partial m_6} & \frac{\partial c_1}{\partial n_1} & \cdots & \frac{\partial c_6}{\partial n_6} & \frac{\partial c_1}{\partial k_1} & \cdots & \frac{\partial c_6}{\partial k_6} & 0 & 0 & 0 & 0 & 0 & 0 \\ 0 & 0 & 0 & 0 & 0 & \cdots & 0 & 0 & \cdots & 0 & 0 & \cdots & 0 & 1 & 0 & 0 & 0 & 0 & 0 \\ 0 & 0 & 0 & 0 & 0 & \cdots & 0 & 0 & \cdots & 0 & 0 & \cdots & 0 & 0 & 1 & 0 & 0 & 0 & 0 \\ 0 & 0 & 0 & 0 & 0 & \cdots & 0 & 0 & \cdots & 0 & 0 & \cdots & 0 & 0 & 0 & 1 & 0 & 0 & 0 \\ 0 & 0 & 0 & 0 & 0 & \cdots & 0 & 0 & \cdots & 0 & 0 & \cdots & 0 & 0 & 0 & 0 & 1 & 0 & 0 \\ 0 & 0 & 0 & 0 & 0 & \cdots & 0 & 0 & \cdots & 0 & 0 & \cdots & 0 & 0 & 0 & 0 & 0 & 1 & 0 \\ 0 & 0 & 0 & 0 & 0 & \cdots & 0 & 0 & \cdots & 0 & 0 & \cdots & 0 & 0 & 0 & 0 & 0 & 0 & 1 \\ 0 & 0 & 0 & 0 & 0 & \cdots & 0 & 0 & \cdots & 0 & 0 & \cdots & 0 & 0 & 0 & 0 & 0 & 0 & 0 \end{pmatrix} \tag{7}$$

Then, Equation (6) can be efficiently addressed as follows:

$$\mathrm{dy} = \mathrm{A}^{-1} \cdot \mathrm{B} \cdot \mathrm{diag}(\mathrm{dz}) \cdot \gamma \tag{8}$$

diag(dz) in Equation (8) above is a diagonal matrix based on vector z, and all the elements of vector γ are 1.

Importantly, the aforementioned equations are applicable strictly under conditions of infinitesimal variable changes. For more accurate decomposition outcomes, it is essential to segment the actual variation in exogenous variables into numerous intervals. Following the original methodology, this study set the interval count $N$ to 16000. The Euler method for numerical integration was employed to calculate the effect of exogenous variables for each interval $N$:

$$\begin{cases} \Theta^{(n)} = \left(A^{(n-1)}\right)^{-1} \cdot B^{(n-1)} \cdot \mathrm{diag}(dz) \\ dy^{(n)} = \Theta^{(n)} \cdot \gamma \\ z^{(n)} = z^{(n-1)} + dz \\ y^{(n)} = y^{(n-1)} + dy^{(n)} \\ A^{(n)} = \lambda\left(y^{(n)}, z^{(n)}\right) \\ B^{(n)} = \omega\left(y^{(n)}, z^{(n)}\right) \end{cases} \quad (9)$$

where $n = 1,2,\cdots,N$ and $dz = \frac{\Delta z}{N}$. Iteratively summing the contributions of each interval yields the desired decomposition result with the expression:

$$\Theta = \sum_{n=1}^{N} \Theta^{(n)} \quad (10)$$

As the exogenous variables $\Delta y_i$ change, the values of operational carbon intensity change within the residential building emissions model produced by the endogenous variables $\Delta z_i$. These changes are what the elements $\theta_{ij}$ of the matrix in Equation (10) represent.

$$\Delta c|_{0\to T} = \Delta p + \Delta g + \Delta h + \Delta e + \Delta m + \Delta n + \Delta k + \Delta s \quad (11)$$

The carbon intensity changes over period *T*, denoted $\Delta c|_{0\to T}$. The expressions in Equation (11) that appear after the equal sign reflect how various drivers contribute to this change. Specifically, Δ*m*, Δ*n*, Δ*k*, and Δ*s* denote the total influence of the end-use structure, the electrification rate, the factor of total emissions relative to electricity-related carbon emissions, and the electricity emission factors across different end uses, respectively. These impacts can be further broken down into the following specific end-use activities:

$$\begin{cases} \Delta m = \Delta m_{Water\ heating} + \Delta m_{Space\ heating} + \Delta m_{Cooking} + \Delta m_{Lighting} + \Delta m_{Space\ cooling} + \Delta m_{Appliances\ \&\ others} \\ \Delta n = \Delta n_{Water\ heating} + \Delta n_{Space\ heating} + \Delta n_{Cooking} + \Delta n_{Lighting} + \Delta n_{Space\ cooling} + \Delta n_{Appliances\ \&\ others} \\ \Delta k = \Delta k_{Water\ heating} + \Delta k_{Space\ heating} + \Delta k_{Cooking} + \Delta k_{Lighting} + \Delta k_{Space\ cooling} + \Delta k_{Appliances\ \&\ others} \\ \Delta s = \Delta s_{Water\ heating} + \Delta s_{Space\ heating} + \Delta s_{Cookting} + \Delta s_{Lighting} + \Delta s_{Space\ cooling} + \Delta s_{Appliances\ \&\ others} \end{cases} \quad (12)$$

Equation (12) illustrates the impact of changes in electrification rates, the factor of total emissions relative to electricity-related carbon emissions, electricity emission factors, and the end-use structure of different household end-use activities on residential building carbon emission intensity.

### *3.3. Decarbonization assessment of building electrification*

The decarbonization intensity of residential building electrification ($DCI$), which is defined as carbon reduction per household, can be assessed by examining the negative impacts of carbon intensity globally on the basis of the DSD decomposition results as follows:

$$DCI|_{0\to T} = -\sum(\Delta c_i|_{0\to T}) \tag{13}$$

where $\Delta c_i|_{0\to T} \leq 0$. Notably, $m$, $n$, and $k$ denote the electrification rate, the factor of total emissions relative to electricity-related carbon emissions, and the emission factors of electricity, respectively. Therefore, the corresponding carbon emission reduction ($DC$) can be calculated through the above decarbonization intensity as follows:

$$DC|_{0\to T} = DCI|_{0\to T} \times H|_{0\to T} \tag{14}$$

*3.4. Data source*

This study collected data from 52 major economies between 2000 and 2021. Population and economic data were obtained from the World Bank (https://data.worldbank.org/indicator), including population (https://data.worldbank.org/indicator/SP.POP.TOTL; see Fig. D1 in Appendix D), GDP (https://data.worldbank.org/indicator/NY.GDP.MKTP.PP.CD; see Fig. D2 in Appendix D), and HFC (https://data.worldbank.org/indicator/NE.CON.PRVT.PP.CD; see Fig. D3 in Appendix D). Among these, HFC and GDP were converted into international dollars at present value using Purchasing Power Parity and were subsequently adjusted using the relevant indices. As shown in Fig. D4, additional data on energy consumption and carbon emissions from end-use activities were sourced from the Global Building Emissions Database (GLOBE, https://globe2060.org/). Furthermore, the number of households was calculated as the ratio of total population to the average household size, with the average household size data obtained from the United Nations Population Division (https://www.un.org/development/desa/pd/data/household-size-and-composition; see Fig. D5 in Appendix D). Additionally, the economies under study were categorized into regions according to their climatic characteristics and socioeconomic factors, as detailed in Appendix B.

## 4. Results

### *4.1. Changes in carbon intensity within operational residential buildings worldwide*

Fig. 2 illustrates the impact of various factors on the carbon intensity of residential buildings during their operational phase. The analysis period spans from 2000 to 2021, corresponding to the data availability. To enhance the clarity and comparability of trends, this study divided the timeline at its midpoint, around 2010, creating two distinct sub-periods: 2000–2011 and 2011–2021. China contributed significantly to the increase in carbon emissions per household, with emission intensity increasing from 121 kilograms of $CO_2$ per household (kg$CO_2$/household) between 2000 and 2011 to 277 kg$CO_2$/household from 2011 to 2021. Furthermore, during these two periods, the United States, Europe (including New Zealand) and Australia presented distinct negative contributions, suggesting that the decrease in household size in these areas helped offset the increase in carbon intensity. The negative effects of GDP per capita on carbon intensity were notable across all countries and regions, particularly in China (5494 kg$CO_2$/household), Australia (5415 kg$CO_2$/household), and the United States (6112 kg$CO_2$/household). This suggests that economic growth led to increased carbon intensity in the operational phase of residential buildings [49]. Household consumption capacity had varying effects on carbon intensity. In 2011–2021, increasing household consumption capacity in China and Morocco significantly led to an increase in carbon intensity, with effects of 466 kg$CO_2$/household and 410 kg$CO_2$/household, respectively. In the United States and Canada, household consumption capacity adversely affected carbon intensity, suggesting regional variations in consumption patterns. On the other hand, energy intensity had a positive effect on changes in carbon intensity across all countries and regions, particularly in the United States and Australia, with effects of 7520 kg$CO_2$/household and 5268 kg$CO_2$/household, respectively. This signifies notable enhancements in energy efficiency, leading to a reduction in carbon intensity for residential buildings [50]. In terms of the rate of electrification, while the effects varied across countries, overall, electrification rates contributed negatively to carbon intensity in 2000–2021, especially in China (3536 kg$CO_2$/household) and Morocco (1558 kg$CO_2$/household), suggesting that higher electrification rates lead to increased carbon emissions to some extent [51]. In contrast, changes in the emission factors of electricity generally made positive contributions worldwide, indicating that advances in electricity generation technologies and the utilization of renewable energy sources have resulted in reduced carbon emissions [52].

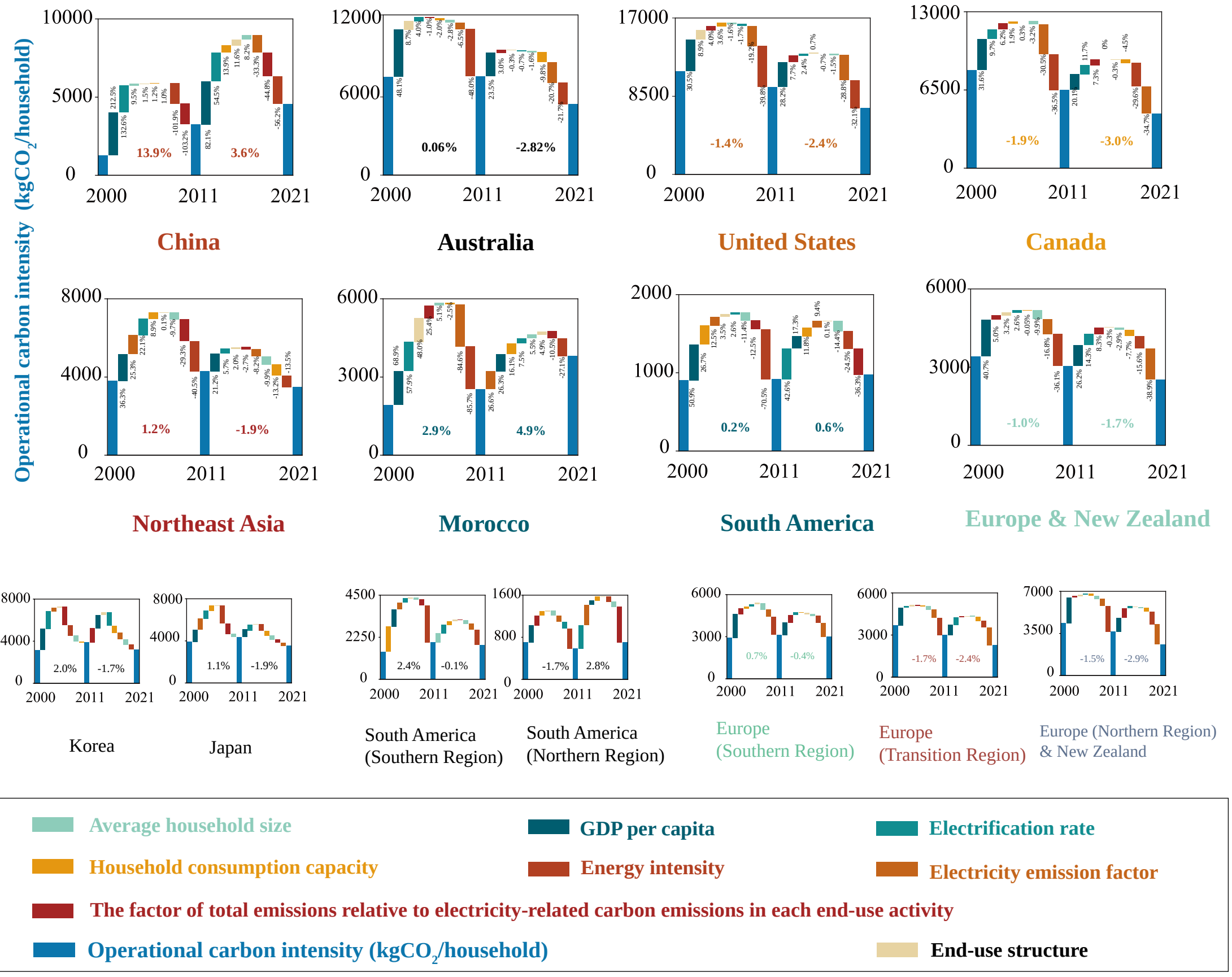

**Fig. 2.** Changes in global residential building operational carbon intensity from 2000 to 2021.

## *4.2. Impacts of the electrification rate and electricity emission factors on carbon intensity*

Figs. 3 a and b illustrate the influence of electrification rates for end-use activities on the carbon emission intensity of residential buildings across different regions and countries. Overall, the electrification rate was a key driver of increased carbon intensity, especially in China, where it accounted for 41.3%, which was significantly higher than that in other countries. Notably, the trends observed in the United States and Australia varied across the two time periods. From 2000 to 2011, the electrification rate in Australia had a negative effect, with an annual average increase of 27.6 $kgCO_2$ per household. In contrast, from 2011 to 2021, the electrification rate (-5.5 $kgCO_2$/household/year) significantly contributed to a reduction in carbon intensity. In the United States, residential electrification led to an increase in carbon intensity from 2000 to 2011, with an average annual rise of 22.3 $kgCO_2$ per household from 2011 onward. In Fig. 3 c, space heating was identified as the primary driver of increased carbon intensity during the operational phase, with an average annual increase of 301 $kgCO_2$ per household. Cooking, water heating, and lighting followed, contributing to increases of 89.0, 17.8, and 13.3 $kgCO_2$/household/year, respectively. In contrast, appliances and others reduced the carbon intensity by 11.4 $kgCO_2$/household/year, whereas space cooling contributed a smaller reduction of 3.1 $kgCO_2$/household/year. An analysis of electrification rates and their effects on carbon intensity at the country level indicated that from 2000 to 2021, space heating in China significantly contributed to carbon intensity, with an increase of 140.5 $kgCO_2$/household/year. In the United States, electrification rates for water heating and space heating posed significant barriers to reducing carbon intensity, with increases of 16.8 $kgCO_2$/household/year and 6.5 $kgCO_2$/household/year, respectively. In contrast, space cooling (-3.1 $kgCO_2$/household/year) and lighting (-19.1 $kgCO_2$/household/year) significantly reduced carbon intensity, reflecting the cleaner electricity supply in the United States [53]. In Australia, space heating (33.3 $kgCO_2$/household/year) and cooking (7.8 $kgCO_2$/household/year) had a more notable negative impact on carbon intensity compared to water heating (-28.9 $kgCO_2$/household/year) and appliances with others (-0.36 $kgCO_2$/household/year). Similarly, the electrification rate of space heating in Canada was also significantly negative on carbon intensity, contributing 73.8 $kgCO_2$/household/year. In most countries, electrification rates of end-use activities negatively impacted carbon intensity. China experienced the most substantial rise in carbon intensity from space heating, increasing from 133 $kgCO_2$/household/year during 2000–2011 to 149 $kgCO_2$/household/year in 2011–2021, indicating a notable upward trend. In the United States, the effects of appliances and others showed an increasing trend in carbon intensity, increasing from -40.0 $kgCO_2$/household/year during 2000–2011 to 3.8 $kgCO_2$/household/year during 2011–2021, indicating that increasing the electrification rate led to increase carbon emissions [54]. Conversely, electrification rates for cooking in the United States (from 0.11 $kgCO_2$/household/year during 2000–2011 to 0.63 $kgCO_2$/household/year during 2011–2021) had a relatively small effect on carbon intensity. In Morocco, the electrification rate of cooking had a significant effect on carbon intensity, especially between 2000 and 2011, when it reached as high as 84.3 $kgCO_2$/household/year. The electrification rates of water heating in Australia and Canada significantly reduced the carbon intensity. Notably, Australia achieved decarbonization in residential water heating electrification, with emissions intensity of -21.5 $kgCO_2$/household/year during the 2011–2021 period.

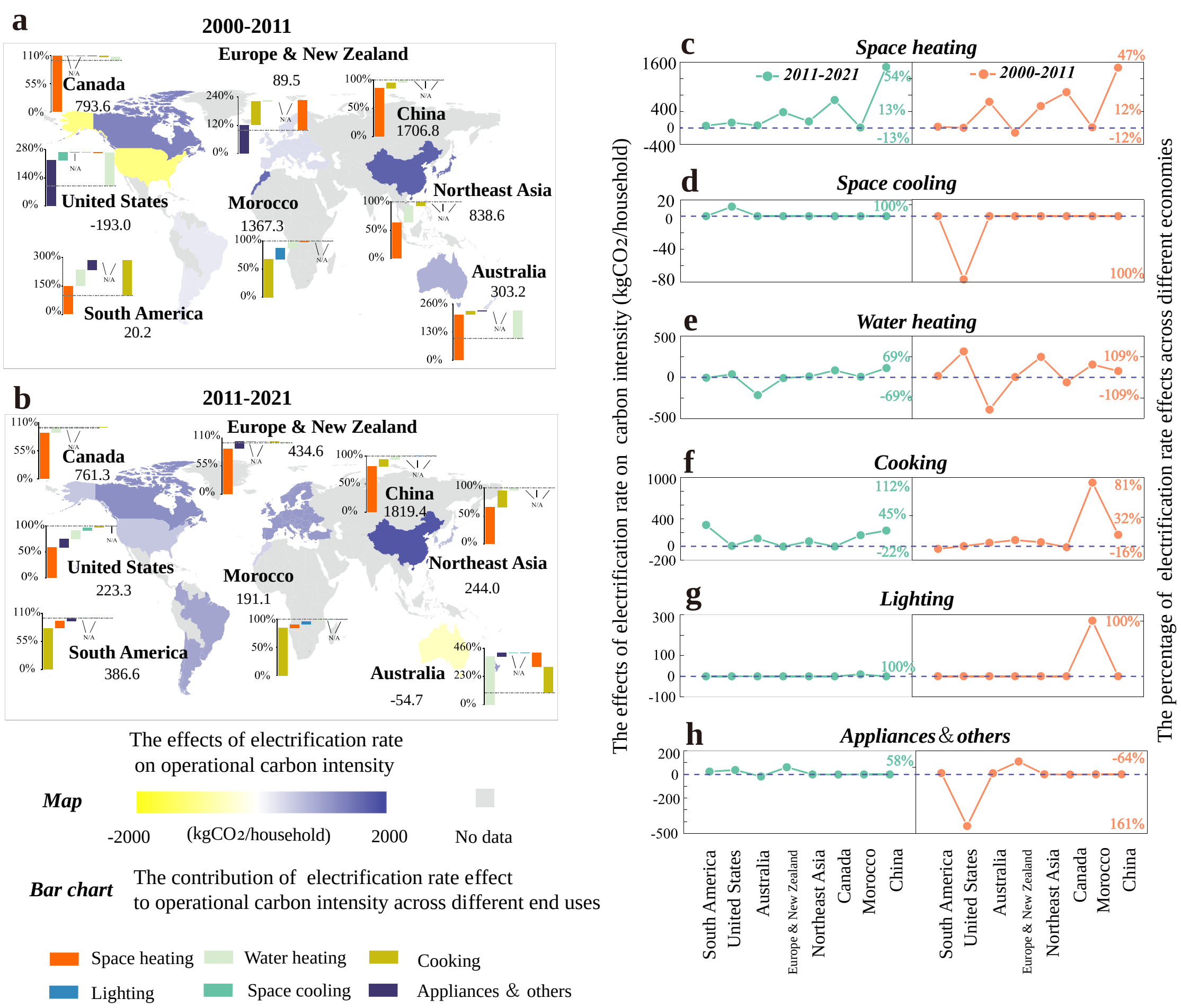

**Fig. 3.** The impact of electrification rates on residential building carbon intensity worldwide: (a-b) total decarbonization impact of residential building electrification rates during 2000–2011 and 2011–2021; (c-h) comparison of the impact of varying end-use electrification rates on residential buildings globally. Note: Considering the data availability, lighting was excluded in Northeast Asia.

Fig. 4 shows the influence of electricity emission factors on carbon intensity. Compared with the influence of the electrification rate, the influence of electricity emission factors on emission reduction was more pronounced. Between 2000 and 2011, developed countries such as Canada (-226 $kgCO_2$/household/year), the United States (-196 $kgCO_2$/household/year), Australia (-44 $kgCO_2$/household/year), and Europe (-52 $kgCO_2$/household/year) achieved significant reductions in carbon intensity. In contrast, China (1.1 $kgCO_2$/household/year) and Northeast Asia (87 $kgCO_2$/household/year) experienced increases, likely due to a greater reliance on fossil fuels in their energy mix during this period. During the 2011–2021 period, Australia (-157 $kgCO_2$/household/year), Canada (-226 $kgCO_2$/household/year), and the United States (-273 $kgCO_2$/household/year) continued to make significant contributions to emission reductions, reflecting the wider adoption of low-carbon technologies and clean energy in power production. Moreover, the effects of electricity emission factors in China significantly decreased (-112 kg $CO_2$/household/year), indicating a reduction in coal dependency and increasing use of renewable energy. Notably, Morocco's effects on electricity emission factors became negative (68 kg$CO_2$/household/year) in the same period, suggesting delays in its energy transition and contributing to rising carbon emissions.

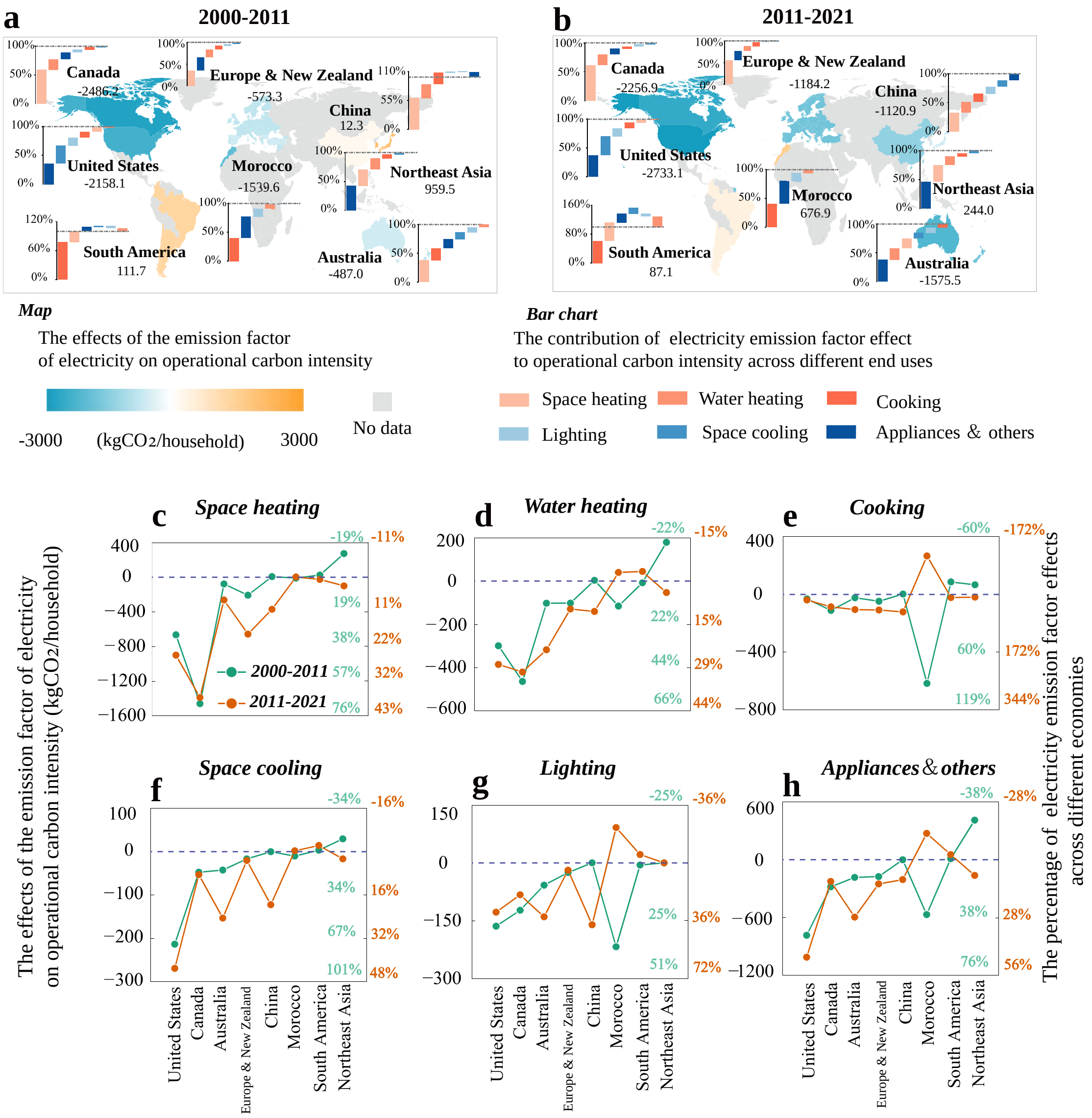

**Fig. 4.** The impact of the emission factors of electricity on carbon intensity: (a-b) total decarbonization impact of residential building electricity emission factors during 2000–2011 and 2011–2021, respectively; (c-h) comparison of the effects of different end-use emission factors of electricity on residential buildings worldwide. Note: Considering the data availability, lighting was excluded in Northeast Asia.

Figs. 4 c and d analyze the impact of electricity emission factors on carbon emission intensity across different countries and regions, comparing the periods 2000–2011 and 2011–2021. In China, from 2000–2011, water heating (0.29 $kgCO_2$/household/year) and cooking (0.24 $kgCO_2$/household/year) significantly increased carbon intensity. However, from 2011 to 2021, the impact of electricity carbon emission factors improved considerably, especially for space heating (from 0.67 to -37 $kgCO_2$/household/year), highlighting China's substantial progress in reducing emissions [55]. In Australia, all end-use activities contributed positively to carbon intensity during 2000–2011, but from 2011 to 2021, significant reductions in emissions were achieved in space heating (-26.3 $kgCO_2$/household/year) and space cooling (-15.3 $kgCO_2$/household/year). Similarly, in the United States, further reductions in appliances and others (from -78.6 during 2000–2011 to -101.3 $kgCO_2$/household/year during 2011–2021) and space heating (from -66.5 during 2000–2011 to -90 $kgCO_2$/household/year during 2011–2021) driven by effective measures were reported. In Canada, emission reductions remained consistent across both periods, with notable reductions in space heating (from -132 $kgCO_2$/household/year during 2000–2011 to -139 $kgCO_2$/household/year during 2011–2021). In Northeast Asia, while high electricity emission factors for space heating and appliances with others were recorded from 2000 to 2011, carbon reduction measures began to take effect between 2011 and 2021, leading to an increase in the contribution of space heating to carbon intensity (-10 $kgCO_2$/household/year). In South America, changes in electricity emission factors from 2011-2021 led to increases in the impacts of space cooling (2.2 $kgCO_2$/household/year) and lighting (5.3 $kgCO_2$/household/year) on carbon intensity. Moreover, Morocco experienced heightened pressure from carbon emissions, particularly from cooking (26.7 $kgCO_2$/household/year). Europe sustained a positive trend in emission reductions during both periods, particularly in space heating (from -18.8 $kgCO_2$/household/year during 2000–2011 to -65.6 $kgCO_2$/household/year during 2011–2021), indicating continuous improvements in the emission factors of electricity reduction efforts across the region.

Sections 4.1 and 4.2 present the global impacts of residential building electrification on carbon intensity, addressing Question 1 posed in Section 1.

## 5. Discussion

### *5.1. Global electrification progress in operational residential buildings*

Figs. 5 a and b depict the variations in electrification rates and electricity emission factors for residential buildings from 2000 to 2021, respectively. During this period, China's electrification rate surged from 28.2% to 48.0%, indicating that it was one of the world's fastest-growing nations in electrification. However, China's emission factors of electricity remained relatively high in the early period, at 5.9 kg$CO_2$/kgce in 2000. Although it gradually decreased, it reached 4.0 kg$CO_2$/kgce by 2021. This indicates that while China has made progress in electrification, further optimization of the power structure is necessary to reduce overall carbon emissions [56]. In contrast, the electrification rates of Australia and the United States remained relatively stable, reaching 51.2% and 47.1%, respectively, in 2021. Australia's electricity emission factor was 8.1 kg$CO_2$/kgce in 2000, with only a modest reduction to 7.4 kg$CO_2$/kgce by 2021. Comparatively, the United States experienced a steady decline in its electricity emission factor, decreasing from 5.7 kg$CO_2$/kgce in 2000 to 3.5 kg$CO_2$/kgce in 2021, demonstrating a consistent improvement in its energy mix. Globally, Europe (including New Zealand) exhibited a significant reduction in the electricity emission factor, which decreased from 3.5 kg$CO_2$/kgce in 2000 to 2.2 kg$CO_2$/kgce in 2021. By 2021, Europe's residential electrification rate consistently increased to 29.5%. In 2021, Morocco's electrification rate reached 23.8%, up from 7.9% in 2000, whereas its electricity emission factor decreased from 12.7 kg$CO_2$/kgce to 6.5 kg $CO_2$/kgce, reflecting significant improvements in its energy structure.

Fig. 5 c depicts the trend of carbon emissions related to electricity use from 2000 to 2021. Global carbon emissions from residential building electricity use rose from 1452 mega tons (Mt) of $CO_2$ to 2032 Mt$CO_2$, reflecting an overall increase of approximately 39.8%. Among these, China's electricity-related carbon emissions showed the highest growth rate, averaging 35.2% annually, indicating a swift rise in electricity demand within the Chinese residential building sector. Countries such as the United States, Australia, and Canada generally maintained stable levels of electricity carbon emissions, with the United States and Canada exhibiting a downward trend. These findings indicate that these countries have made significant progress in building energy efficiency and implementing effective energy policies [57]. Carbon emissions in Northeast Asia increased during the 2000–2013 period, but the growth rate was noticeably lower than that in China. Carbon emissions from electricity use in South America and Morocco were relatively low but showed a clear upward trend since 2010, which is likely related to economic development and increased energy demand. Europe's carbon emissions from electricity use showed a downward trend since 2010, reflecting its strong commitment to energy transition policies and the promotion of green building initiatives.

In summary, the trends of residential building carbon emissions from electricity use, electrification rates and electricity emission factors answer Question 2 in Section 1.

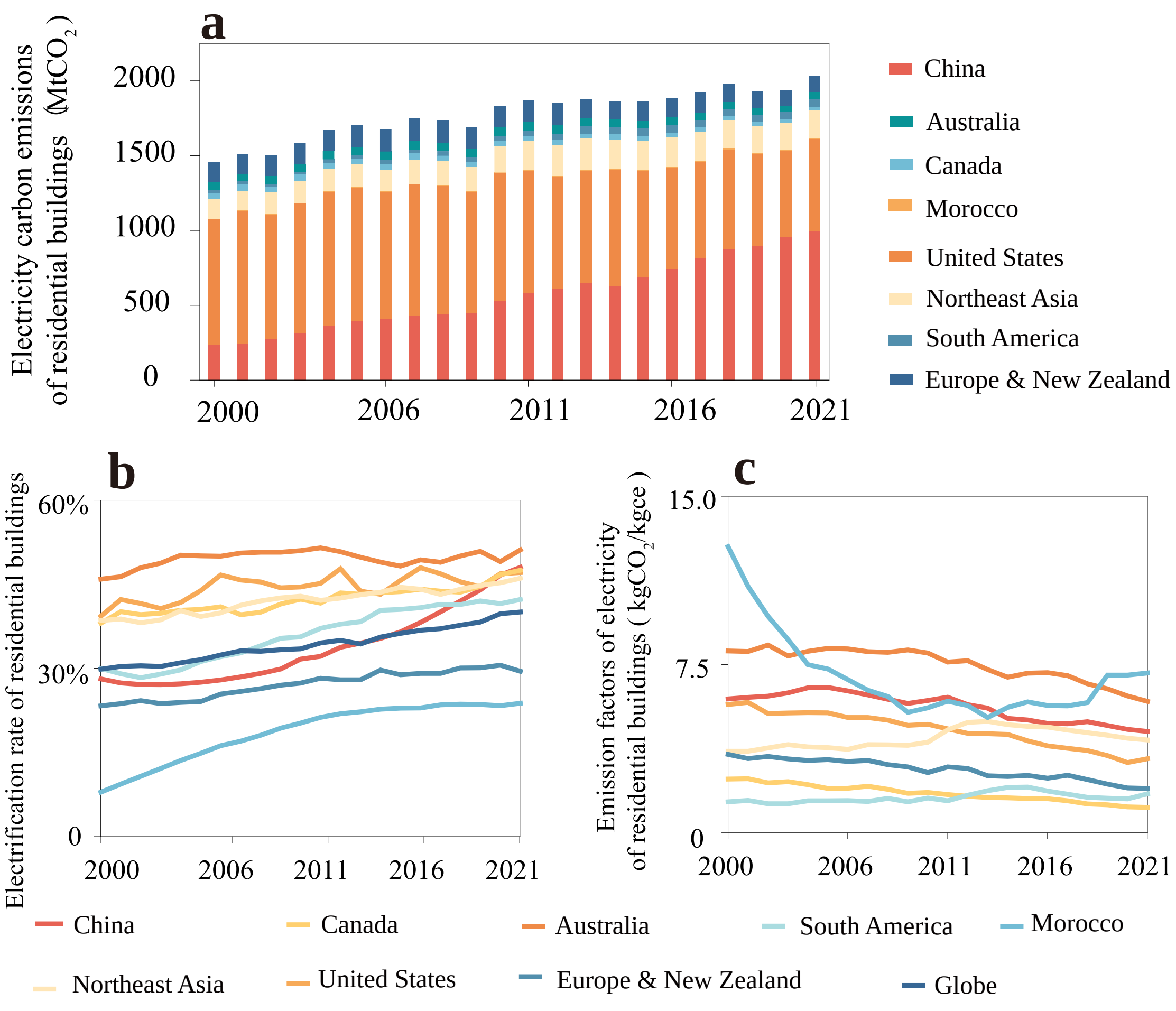

**Fig. 5.** Global electrification progress in residential building operations: (a) carbon emissions related to electricity use in operational residential buildings from 2000 to 2021; (b) global trends in residential building electrification rates from 2000 to 2021; (c) changes in electricity emission factors of residential buildings worldwide from 2000 to 2021.

*5.2. Historical decarbonization of end-use electrification worldwide*

Fig. 6 shows an analysis of decarbonization metrics across various countries and regions during the operational phase. These metrics include total decarbonization, decarbonization per capita, and decarbonization per household. As shown in Fig. 6 a, China's decarbonization fluctuated significantly between 2001 and 2021, increasing from 54.1 $MtCO_2$ in 2001 to 235 $MtCO_2$ in 2020. In contrast, Australia and the United States achieved decarbonization of 3.1 $MtCO_2$ and 48.3 $MtCO_2$, respectively, in 2021. In 2002, the United States peaked decarbonization of 183 $MtCO_2$, far exceeding that of other countries, but subsequently experienced a steady decline to 48.3 $MtCO_2$ by 2021. This trend may be attributed to building energy efficiency policies and technological innovations during that period. Canada's decarbonization remained relatively stable, peaking at 106 $MtCO_2$ in 2015. Moreover, Europe experienced a gradual increase in decarbonization, reaching 23.1 $MtCO_2$ in 2020. Fig. 6 b shows that China's decarbonization per household rose sharply from 170 $kgCO_2$/household in 2001 to 744 $kgCO_2$/household in 2020, possibly because of the growing number of residential buildings [58]. Australia's decarbonization per household peaked at 436 $kgCO_2$/household in 2011, fluctuating thereafter to 305 $kgCO_2$/household in 2021, reflecting relatively stable electrification levels. The United States recorded a decarbonization per household of 1661 $kgCO_2$/household in 2002. Fig. 6 c shows that China's decarbonization per capita rose from 42.9 $kgCO_2$ per capita in 2001 to 49.2 $kgCO_2$ per capita in 2021. In Australia, decarbonization per capita reached 123 $kgCO_2$ per capita in 2021, an increase from 30.4 $kgCO_2$ per capita in 2001. While the increase was significant, the growth was more moderate, reflecting the gradual pace of its electrification process. In the United States, decarbonization per capita stood at 146 $kgCO_2$ in 2021, whereas it stood at 109 $kgCO_2$ per capita in 2001. In Europe, decarbonization per capita reached 113 $kgCO_2$ in 2021, reflecting stable electrification levels and a strong focus on sustainable development. This highlights differences in residential electrification levels and policy directions between different countries [59].

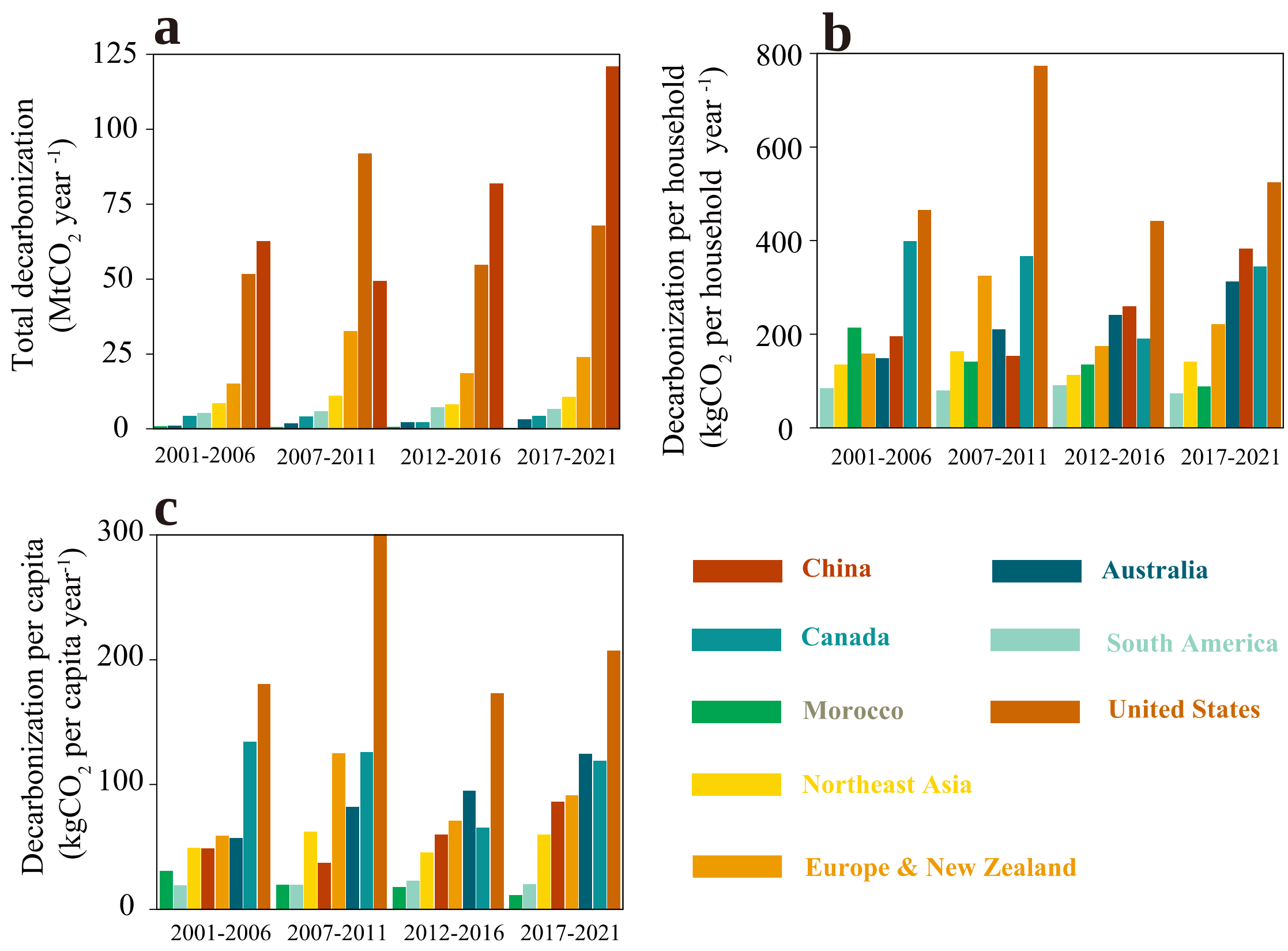

**Fig. 6.** Decarbonization effects of electrification on residential buildings worldwide: (a) comparison of average annual decarbonization from 2001 to 2021; (b) comparison of average annual decarbonization intensity during the phases of 2001–2006, 2007–2011, 2012–2016 and 2017–2021; (c) comparison of average annual decarbonization per capita over the periods of 2001–2006, 2007–2011, 2012–2016 and 2017–2021.

As shown in panels a-d of Fig. 7, global residential electrification has cumulatively achieved a decarbonization effect of 3954 $MtCO_2$ from 2001 to 2021. Figs. 7 e and f further illustrate the decarbonization contributions of different end-use electrification processes. Among these, the electrification of space heating in residential buildings was the primary contributor, with its proportional contribution varying across different periods (2001–2006, 2007–2011, 2012–2016, 2017–2021): 65.7%, 62.4%, 42.7%, and 66.5%, respectively. This was followed by water heating, appliances with others, and cooking. The decarbonization effects from space cooling and lighting were relatively similar, with comparable proportions across all periods. The assessment of decarbonization through the electrification of space heating indicates that China achieved a significant reduction of 480 $MtCO_2$ from 2017 to 2021, followed by a reduction of 299 $MtCO_2$ during the 2001–2006 period. In contrast, the decarbonization contributions during the 2007–2011 and 2012–2016 periods were relatively similar, at 159 $MtCO_2$ and 155 $MtCO_2$, respectively. In the United States, the impact of decarbonization through space heating electrification fluctuated across different periods. The decarbonization amount was 293 $MtCO_2$ during 2007–2011, followed by 196 $MtCO_2$ from 2017 to 2021. Moreover, the reductions in the 2001–2006 and 2012–2016 periods were 181 $MtCO_2$ and 135 $MtCO_2$, respectively. Other countries and regions, such as Australia, Canada, and Northeast Asia, showed growth in their contributions to decarbonization through the electrification of space heating, but their overall decarbonizations remained relatively modest. In terms of decarbonization through the electrification of appliances and others, from 2001 to 2016, the United States dominated with a decarbonization of 55.9 $MtCO_2$, whereas China (4.2 $MtCO_2$) and Europe (12.0 $MtCO_2$) showed growth potential, with other regions progressing slowly. From 2007 to 2011, China's decarbonization increased to 16.7 $MtCO_2$. The United States (101.2 $MtCO_2$) maintained its global leadership, whereas Europe (24.7 $MtCO_2$) experienced significant growth. South America's decarbonization rose to 4.1 $MtCO_2$, reflecting the early adoption of renewable energy. Between 2012 and 2016, China's decarbonization increased to 45.2 $MtCO_2$, the United States' decarbonization decreased to 74.2 $MtCO_2$, and Europe's decarbonization remained stable at 19.1 $MtCO_2$. From 2017 to 2021, China's decarbonization decreased to 25.3 $MtCO_2$. The United States recovered to 76.7 $MtCO_2$, and Europe slightly increased to 19.9 $MtCO_2$. Northeast Asia and South America experienced significant growth. The carbon reduction from the electrification of water heating in China showed significant fluctuations, peaking at 122.6 $MtCO_2$ during 2012–2016 and before decreasing to 43.8 $MtCO_2$ from 2017 to 2021. These variations may reflect changes in residential water heating demand or the impact of equipment upgrades. In contrast, decarbonizations in the United States, Australia, and Europe remained relatively stable, with no significant fluctuations observed. In terms of cooking electrification, China's carbon reduction gradually declined, from 20.5 $MtCO_2$ in the 2000–2006 period to 14.1 $MtCO_2$ from 2017 to 2021. Notably, Morocco achieved a reduction of 2.2 $MtCO_2$ through cooking electrification during 2001–2006, whereas South America achieved a reduction of 19.6 $MtCO_2$, demonstrating the potential of emerging economies in cooking electrification.

In summary, significant differences existed in the carbon reduction contributions from electrification across various end-use activities in different countries. Both China and the United States made significant strides in reducing carbon emissions through the electrification of various end uses, while Europe also contributed notably to areas such as space heating and appliances with others. The decarbonization effects of these end-use activities exhibited phase fluctuations, with space heating and water heating electrification demonstrating substantial reduction potential at all stages, whereas lighting and space cooling electrification remained more limited.

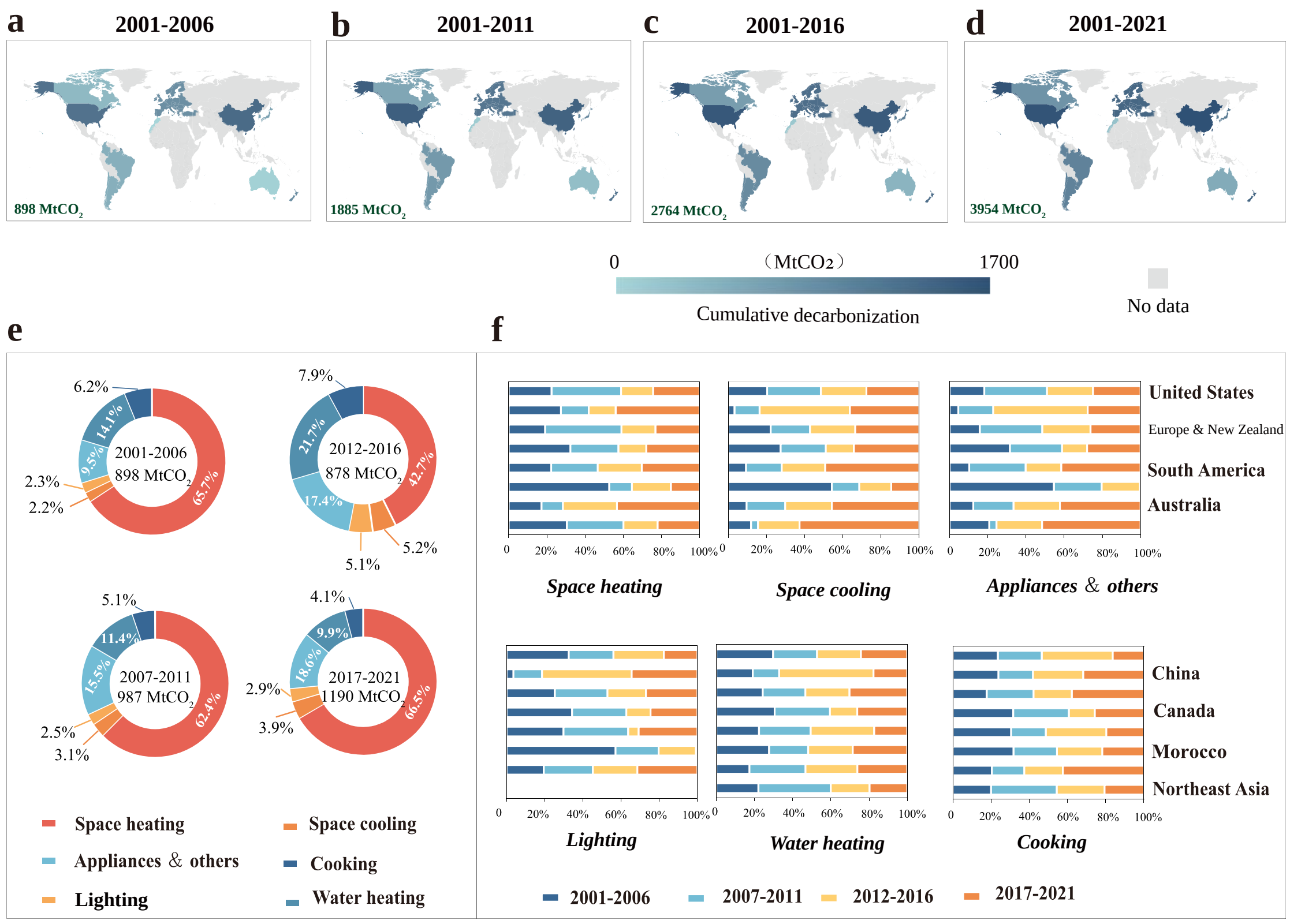

**Fig. 7.** The decarbonization impacts of electrification on residential buildings worldwide: (a-d) the spatiotemporal evolution of the decarbonization effect of residential building electrification worldwide; (e) the decarbonization effect of electrification across different end-use activities across the phases of 2001–2006, 2007–2011, 2012–2016, and 2017–2021; (f) percentage of the decarbonization effect from different end-use electrification across four phases (2001–2006, 2007–2011, 2012–2016 and 2017–2021) in different countries. Note: Considering the data availability, lighting was excluded in Northeast Asia.

### *5.3. Policy implications*

According to the above analysis, increasing the residential electrification rate does not always reduce carbon emissions, as electrification itself is not equivalent to energy decarbonization. If the power supply still relies primarily on fossil fuels, increasing the electrification rate may merely shift carbon emissions from traditional energy sources to the electricity production sector without achieving an overall reduction in emissions. Therefore, simply increasing the electrification rate is not a fundamental solution to the carbon emission issue. Only under the premise of a clean and efficient energy structure can improvements in electrification genuinely lead to a reduction in carbon emissions. On this basis, the following strategies are proposed:

**(a) Implementing policy and market mechanisms.** To encourage the low-carbon transition of residential electrification, financial subsidies and tax incentives can be provided to homeowners [27]. Financial subsidies can mitigate initial expenses for energy-efficient installations [60, 61], and tax incentives can reduce the costs of home ownership while promoting low-carbon technology adoption [62, 63]. These measures can help promote the use of clean energy sources [64], lower greenhouse gas emissions and support the development of a building sector that prioritizes sustainability and environmental friendliness [65].

**(b) Low-carbon electrification**. The development of renewable energy aims to promote low-carbon power production [66] and accelerate the decarbonization process. First, efforts should focus on transforming and upgrading thermal power units by coordinating coal power energy-saving measures, reducing consumption [67], increasing heating and flexibility [68], and increasing clean energy usage to reduce thermal coal dependence. Second, the expansion of renewable energy capacity should be prioritized to replace thermal power generation units [69]. Finally, promoting the construction of energy storage facilities will enable peak and valley electricity adjustments and other scenarios, thereby enhancing the comprehensive utilization of renewable energy [70].

**(c) Demand side management**. Reasonable control of the total amount and intensity of electricity consumption promotes decreased electricity demand in the building sector [71]. First, we should decrease building electricity intensity; enhance energy-saving design standards for new constructions; encourage the development of ultralow, near-zero energy, and zero-carbon buildings; and prevent unnecessary electricity use [72]. Second, improving the energy efficiency of electrical appliances and advocating for energy-efficient products are vital for fulfilling the energy requirements of building heating, hot water, air conditioning, cooking, and lighting. This will also optimize electricity consumption in buildings [73]. Finally, we should promote behavioral energy conservation, implement an energy consumption quota system for residential buildings, guide high energy consumption residential buildings to implement adjustments or renovations, optimize residential ladder electricity price policies, and advocate rational electricity consumption for residents [74].

**(d) Carbon sink management**. First, building design should be optimized by incorporating green roofs and walls to increase vegetation coverage, which enhances natural cooling, reduces reliance on air conditioning, and lowers carbon emissions [75]. Second, green vegetation, such as trees, lawns, and gardens, is planted to absorb carbon dioxide through photosynthesis and release oxygen, creating natural carbon sinks [76]. Third, the utilization of sustainable materials should be promoted to lower the carbon footprint throughout the building's lifecycle [77]. Residents should be encouraged to participate in greening activities, such as home gardens, to increase the community's carbon sink capacity. Finally, energy efficiency can be improved by installing energy-saving appliances and lighting systems, reducing energy consumption, and indirectly decreasing carbon emissions [78].

Overall, Sections 5.2 and 5.3 evaluate the historical progress of decarbonization through end-use electrification and suggest specific strategies to speed up the global shift toward electrifying residential buildings, addressing Question 3 outlined in Section 1.

## 6. Conclusion

This study employed the DSD approach to investigate changes in the carbon intensity of residential buildings globally, emphasizing the importance of electrification for achieving decarbonization. Using this model, this study assessed and compared the role of residential building electrification in reducing carbon emissions during building operations around the world and explored the potential ability of electrification to reduce emissions from various end-use activities in detail. In addition, based on the actual impact of current residential building electrification on decarbonization, corresponding decarbonization strategies and suggestions were proposed. The most important findings of the study are as follows:

### *6.1. Key findings*

- **The increase in operational carbon intensity was due mainly to increased electrification, whereas changes in electricity emission factors positively impacted global decarbonization.** From 2000 to 2021, space heating significantly contributed to increasing carbon intensity, with China's residential heating demand experiencing the most rapid growth, increasing by 140 $kgCO_2$/household annually. In the United States, water heating and space heating contributed to increases in carbon intensity, whereas space cooling and lighting contributed to decreases. Space heating and cooking in Australia significantly increased carbon intensity, but the electrification of water heating successfully reduced emissions. The electrification of space cooling in Morocco significantly increased carbon intensity, especially in 2000–2011. Between 2000 and 2021, the electricity emission factor notably influenced the reduction in carbon intensity in developed regions such as Canada, Australia, the United States and Europe. The impact of the electricity emission factor in China turned positive from 2011 to 2021, indicating increased use of renewable energy, whereas the effects of the electricity emission factor in Morocco turned negative during the same period.
- **The global electrification has increased, with the electrification rate increasing from 29.9% in 2000 to 40.1% in 2021. However, electrification progress has varied across economies.** From 2000 to 2021, the total global carbon emissions from residential building electricity use showed an overall upward trend, increasing from 1452 $MtCO_2$ to 2032 $MtCO_2$, an increase of approximately 39.8%. China has experienced the most significant growth in electricity carbon emissions, whereas emissions in other countries have decreased. From 2000 to 2021, China's residential building electrification rate increased significantly, from 28.2% to 48.0%. The emission factor of electricity remained elevated, necessitating further optimization of the electricity structure. While electrification rates in Australia and the United States were relatively stable, Australia's electricity emission factor remained high, whereas it was significantly lower in the United States. Europe experienced notable advancements in both electrification rates and electricity emission factors, whereas Morocco achieved a significant reduction in its electricity emission factor despite having a low electrification rate.

- **The decarbonization effect of electrification on end-use activities in global residential buildings from 2001 to 2021 contributed to a cumulative reduction of 3954 Mt$CO_2$.** Space heating electrification was the main contributor to decarbonization. For example, China reduced 480 Mt$CO_2$ from 2017–2021 and 299 Mt$CO_2$ from 2001–2006 and achieved similar reductions during 2007–2011 and 2012–2016, at 159 Mt$CO_2$ and 155 Mt$CO_2$, respectively. Between 2001 and 2021, the United States (308 Mt$CO_2$) achieved the greatest decarbonization in the electrification of appliances and others, China and Europe demonstrated steady progress. From 2001 to 2021, China and South America were the main contributing regions to the decarbonization of cooking electrification, accounting for 87 Mt$CO_2$ and 64 Mt$CO_2$, respectively. The carbon reduction resulting from water heating electrification fluctuated significantly in China but remained stable in the United States, Australia, and Europe. The carbon reduction contributions from the electrification of space cooling and lighting were generally limited.

### *6.2. Limitations and future work*

This study has several limitations that should be acknowledged. First, it does not account for cross-country variations in climate conditions—such as differences in heating and cooling degree days (HDD and CDD)—which may introduce biases in the estimated contributions of electrification to decarbonization. This limitation could lead to an overestimation of benefits in warmer climates and an underestimation in colder regions, particularly for temperature-sensitive end uses such as space heating and cooling. Second, due to the limited availability and consistency of high-resolution, long-term national data—especially for developing countries—some socioeconomic and technological factors were analyzed at an aggregated level. Additionally, country-specific electricity generation mixes were not fully incorporated, which may obscure important local differences.

Future research should aim to integrate climate-adjusted thermal loads to more accurately capture regional variations in heating and cooling demands. Incorporating standardized indicators such as HDD and CDD would enhance the precision of end-use-specific assessments. Moreover, improved data availability on national electricity generation structures, household energy behaviors, appliance adoption patterns, and building characteristics would support more robust estimations of emissions reductions attributable to electrification. These improvements would enable deeper insights into the drivers and impacts of residential electrification and inform the development of more targeted and effective decarbonization strategies.

## Appendix

Please find the appendix in the supplemental materials (e-component) related to this submission.

## Acknowledgment

The first author appreciates the National Planning Office of Philosophy and Social Science Foundation of China (24BJY129).